\documentclass[%
reprint,
 amsmath,amssymb,
 aps,
]{revtex4-2}

\usepackage{graphicx}
\usepackage{dcolumn}
\usepackage{bm}
\usepackage{xcolor}

\usepackage{hyperref}

\begin{document}

\preprint{APS/123-QED}

\title{Configurational Entropy of Self Propelled Glass Formers}

\author{Sachin C N and Ashwin Joy}
 \email{ashwin@physics.iitm.ac.in}
\affiliation{%
	Department of Physics, Indian Institute of Technology Madras, Chennai, Tamil Nadu 600036, India 
}%

\date{\today}

\begin{abstract}
  The configurational entropy is an indispensable tool to describe super-cooled liquids near the glass transition. Its calculation requires the enumeration of the basins in the potential energy landscape and when available, it reveals a direct connection with the relaxation time of the liquid. While there are several reports on the measurement of configurational entropy in passive liquids, very little is understood about its role in active liquids which have a propensity to undergo a glass transition at low temperatures. In this paper, we report a careful calculation of the configurational entropy in a model glass former where the constituent units are self propelled. We show that unlike passive liquids, the anharmonic contribution to the glass entropy in these self-propelled liquids can be of the same order as the harmonic contribution, and therefore must be included in the calculation of the configurational entropy. Our extracted configurational entropy is in good agreement with the generalized Adam-Gibbs relation predicted by the random first order transition theory enabling us to deduce a scaling relation between the configurational entropy and the point-to-set length scale in these active systems. Our findings could be of great utility in conventional active systems such as self-propelled granules, Janus particles and dense bacterial suspensions, to mention a few.

\end{abstract}

\maketitle

\section{Introduction}
Active matter systems remain non-equilibrium due to energy conversion at the level of constituent particles that eventually leads to a systematic motion. They span a wide range of physical systems \textemdash from bacterial suspensions \cite{PhysRevLett.93.098103, PhysRevLett.107.028102} and ant colonies \cite{gravish2015glass} to even bird flocks \cite{cavagna2014bird} and fish schools \cite{hubbard2004model}. The constituent units have their size ranging from a few micrometer to a meter. The effect of self-propulsion results in a plethora of exotic collective behavior whose counterpart is absent in the passive systems \cite{LBerthier, gonzalez2012soft, Theurkaff, PhysRevLett.109.248109, bricard2015emergent, klongvessa2019active}. Some of these effects can also be witnessed in jamming \cite{bechinger2016active}, intermittency \cite{mandal2020extreme}, non-trivial velocity correlations \cite{henkes2020dense}, phase separation \cite{PhysRevLett.110.055701,caprini2020spontaneous,fily2014freezing}, active liquid meta-materials \cite{vitelli-natphys} and even phase transitions \cite{czirok1999collective, fily2012athermal}. Experimental studies on the living systems have revealed intriguing behavior which resembles the ubiquitous glass transition. For e.g, in the experiment of zebrafish embryonic explants \cite{schoetz2013glassy}, the formation of a cage by the neighboring particles results in short time sub-diffusive behavior that is reminiscent of glassy dynamics. This makes active matter a new paradigm to investigate fundamental issues  in glass transition. One of the most debated question in glass physics is how the relaxation time scales diverge as the liquid approaches glass transition temperature? \cite{egami2020range, debenedetti2001supercooled}. The random first order transition (RFOT) theory suggests that the relaxation of the liquid occurs through the rearrangement of the correlated mosaics of local domains which are not periodic \textemdash the mosaics being the region of unique configurations \cite{lubchenko2007theory, kirkpatrick1987connections, PhysRevA.40.1045}. It also suggests that the driving force for the rearrangements of these correlated regions is the configurational entropy $S_{\text{c}}$. The extension of RFOT theory on active matter \cite{nandi2018random, mandal2022random} has revealed that the effect of activity on the glassy behavior can be elucidated by studying the microscopic ways in which activity influences $S_{\text{c}}$. This has resolved the apparent diverging predictions that the effect of activity can either enhance \cite{flenner2016nonequilibrium} or mitigate \cite{mandal2016active} the glassy behavior. Though the configurational entropy plays a crucial role in explaining the effect of self-propulsion on the glassy behavior, the detailed simulation studies of the same is sparse in the literature for self-propelled glass formers. While there have been some effort \cite{preisler2016configurational} to compute $S_{\text{c}}$ of an active system using an effective temperature approach, its connection with the relaxation dynamics and the size of the cooperatively rearranging regions especially in the vicinity of glass transition, remains to be explored. This is a pressing issue that requires a careful examination of the 
potential energy landscape (PEL) and its connection to $S_{\text{c}}$. We aim to bridge this gap by providing a detailed study of how activity influences the movement of the system on this landscape, in turn affecting the configuration entropy $S_{\text{c}}$ and therefore the relaxation dynamics. To achieve this, we start with configurations equilibrated at some effective temperature and self propulsion as input to the energy minimization, to find the local minima of the PEL \textemdash referred to as the inherent structures (IS). 
The multiplicity of these IS directly lead us to an estimate of the configurational entropy that is in very good agreement with the generalized Adam-Gibbs (AG) relationship predicted by the RFOT. Our work reveals that unlike passive liquids, the anharmonic contribution to the glass entropy in these self-propelled liquids can be of the same order as the harmonic contribution, and therefore must be included in any calculation of $S_{\text{c}}$. Following this we extract a point-to-set (PTS) length scale $\xi_{\text{s}}$ by randomly pinning a fraction of particles in an equilibrated configuration. Finally by invoking a generalized Adam-Gibbs relation, we derive a scaling relation between $S_{\text{c}}$ and $\xi_{\text{s}}$. We believe these results could be very useful in physical systems such as self-propelled granular systems \cite{deseigne2010collective} and active Janus colloidal particles \cite{ginot2015nonequilibrium} to name a few. The organization of this paper is as follows, in section \ref{SimMod-section} we describe the simulation model used in this study, section \ref{ConfEnt-section} contains the detailed information about the calculation of the configurational entropy, in section \ref{LenScale-section} we have explained the extraction of length scale and its connection to the configurational entropy through RFOT. Finally, in section \ref{conc-section}, we provide the conclusions of our results.
\section{Simulation Model}
\label{SimMod-section}
In this study, we use a numerical model in which the motion of each active particle is governed by the Ornstein-Uhlenbeck (OU) process. In the overdamped limit, equations of motion for the $\text{i}^{\text{th}}$ particle is given by
\begin{align}
	\bm{\dot{r}}_i &=   \frac{1}{m\gamma} \biggl(- \nabla_i U + \bm{f}_i\biggr) \nonumber \\
	\bm{\dot{f}}_i &= \frac{1}{\tau_{\text{p}}} \biggl(- \bm{f}_i + \sqrt{2 m \gamma k_B T_{\text{eff}}} \;\bm{\eta}_i \biggr)
	\label{AOUP}
\end{align}
where $U$ is the interaction potential between the particles in the system that is described later and $\gamma$ is the friction coefficient. This active OU stochastic process  is an athermal model as it lacks the thermal noise and has been extensively used in the studies of athermal active fluids \cite{koumakis2014directed,PhysRevE.91.062304,marconi2015towards}. We drop the hydrodynamic interactions in our model as they are decoupled from the glass transition that happens over a much longer time scale. In this model, activity is therefore completely controlled by two parameters, namely, the effective temperature $T_{\text{eff}}$ which sets the strength of the activity and a time scale $\tau_{\text{p}}$, which sets the duration of persistent motion. The active force $\bm{f}_i$  has the following time correlation.
\begin{equation}
\left< \bm{f(t)} \cdot \bm{f(0)} \right> = \frac{T_{\text{eff}}}{\tau_{\text{p}}} \;\text{exp}(-t/\tau_{\text{p}}) 
\end{equation}
 The Gaussian white noise has the following statistics-
 \begin{equation}
   \left<  \eta^\chi_i (t)\right> = 0, \;\left<  \eta^{\chi}_i(t) \eta^{\zeta}_j(t') \right> = \delta_{\chi\zeta}\delta_{ij}\delta(t - t')
 \end{equation}
 Here $\chi, \zeta$ represent spatial indices and $i,j$ denote particle labels. In the limit of $\tau_{\text{p}} \rightarrow 0$, our model described by eq. \ref{AOUP} reduces to the standard overdamped Brownian dynamics that also satisfies the detailed balance \cite{PhysRevLett.117.038103}. Here the mass of the each particle $m$, the Boltzmann constant $k_B$ and the friction coefficient $\gamma$ are all set to unity in all the production runs. The simulation details of our work are as follows.  
The equations of motion in Eq. \ref{AOUP} is integrated using a fully implicit scheme with a time step of $2 \times 10^{-4}$ which confirms the numerical stability of the production runs throughout the parameter space \cite{PhysRevA.40.3381}. We have employed the Kob-Andersen model in the simulation which is a binary glass forming liquid with $80\%$ large $(L)$ and $20\%$ small $(S)$ particles interacting via the Lennard-Jones (LJ) potential energy \cite{KOBANDERSONLJPOTENTIAL} at
density $\rho = 1.2$. To achieve speed up, we truncated the interaction potential $\phi$ and its two derivatives at a cut-off distance of $r_c$.
\begin{widetext}
\begin{equation}
  U = \frac{1}{2} \sum_{i\neq j} \phi_{ij};\quad \phi_{ij} = 
  \begin{cases}
    4 \epsilon_{ij} \bigg[ \bigg(\dfrac{\sigma_{ij}}{r_{ij}} \bigg)^{12} - \bigg(\dfrac{\sigma_{ij}}{r_{ij}} \bigg)^{6}\bigg], & 0 < r_{ij} \leq r_m\\
    \epsilon_{ij} \bigg[ A\bigg(\dfrac{\sigma_{ij}}{r_{ij}} \bigg)^{12} - B\bigg(\dfrac{\sigma_{ij}}{r_{ij}} \bigg)^{6} + \sum_{n=0}^3 C_{2n} \bigg(\dfrac{r_{ij}}{\sigma_{ij}} \bigg)^{2n}\bigg], & r_m < r_{ij} \leq r_c\\
    0, & r_{ij} > r_c
  \end{cases}
\label{KA}
\end{equation}
\end{widetext}

\noindent The constants $A$, $B$, $C_0$, $C_2$, $C_4$ and $C_6$ are fixed by matching the potential and its two derivatives at the boundaries $r_m = 2^{1/6} \sigma_{ij}$ and the $r_c = 2.5\sigma_{ij}$. The units of length, energy and time are respectively taken as $\epsilon_{LL}$, $\sigma_{LL}$ and $\sqrt{m \sigma_{LL}^2 /\epsilon_{LL}}$. In these units, the values of the potential parameters are set  as, $\epsilon_{LL} = 1.0$, $\epsilon_{SS} = 0.50$, $\epsilon_{LS} = 1.50$, $\sigma_{LL} = 1.0$, $\sigma_{SS} = 0.8$. All the simulations were carried out at a fixed volume in a cubic box with $1000$ particles. The length of each side in reduced units is taken as $9.41036$ and we have used periodic boundary conditions in all the three directions. The range of effective temperatures explored in this paper are $0.45 \leq T_{\text{eff}} \leq 0.9$ and the range of persistence times covered are over three orders of magnitude $2\times 10^{-4} \leq \tau_{\text{p}} \leq 0.1$. In the following section, we give the details of our calculations of the configurational entropy in this active system.

\section{Configurational Entropy}
\label{ConfEnt-section}
Conventional phase transitions such as a liquid to crystal transition is accompanied by the emergence of a long range structural order. A super-cooled liquid does not reveal such long range ordering near the glass transition even though its relaxation time increases by several orders of magnitude. This happens because the number of available distinct states in the PEL becomes sub-extensive as the liquid approaches glass transition. The enumeration of these distinct states therefore directly leads to the estimation of $S_{\text{c}}$. In practice, this is evaluated as the difference between the total and the glass entropy:
\begin{equation}
        S_{\text{c}} = S_{\text{tot}} - S_{\text{glass}}
        \label{S-conf-eq}
\end{equation}
where the total entropy $S_{\text{tot}} = S_{\text{id}} + S_{\text{exc}}$, is calculated by a thermodynamic integration \cite{banerjee2014role,berthier2019configurational}. The glass entropy $S_{\text{glass}}$ carries the contribution from the vibrational degrees of freedom and is  determined from the PEL. The input to all these calculations are the snapshots at some steady state that is characterized by an effective temperature and a persistence time. These calculations are discussed further in the following sections.

\subsection{Total entropy $S_{\text{tot}}$}
\begin{figure}
 	\includegraphics[scale=0.4]{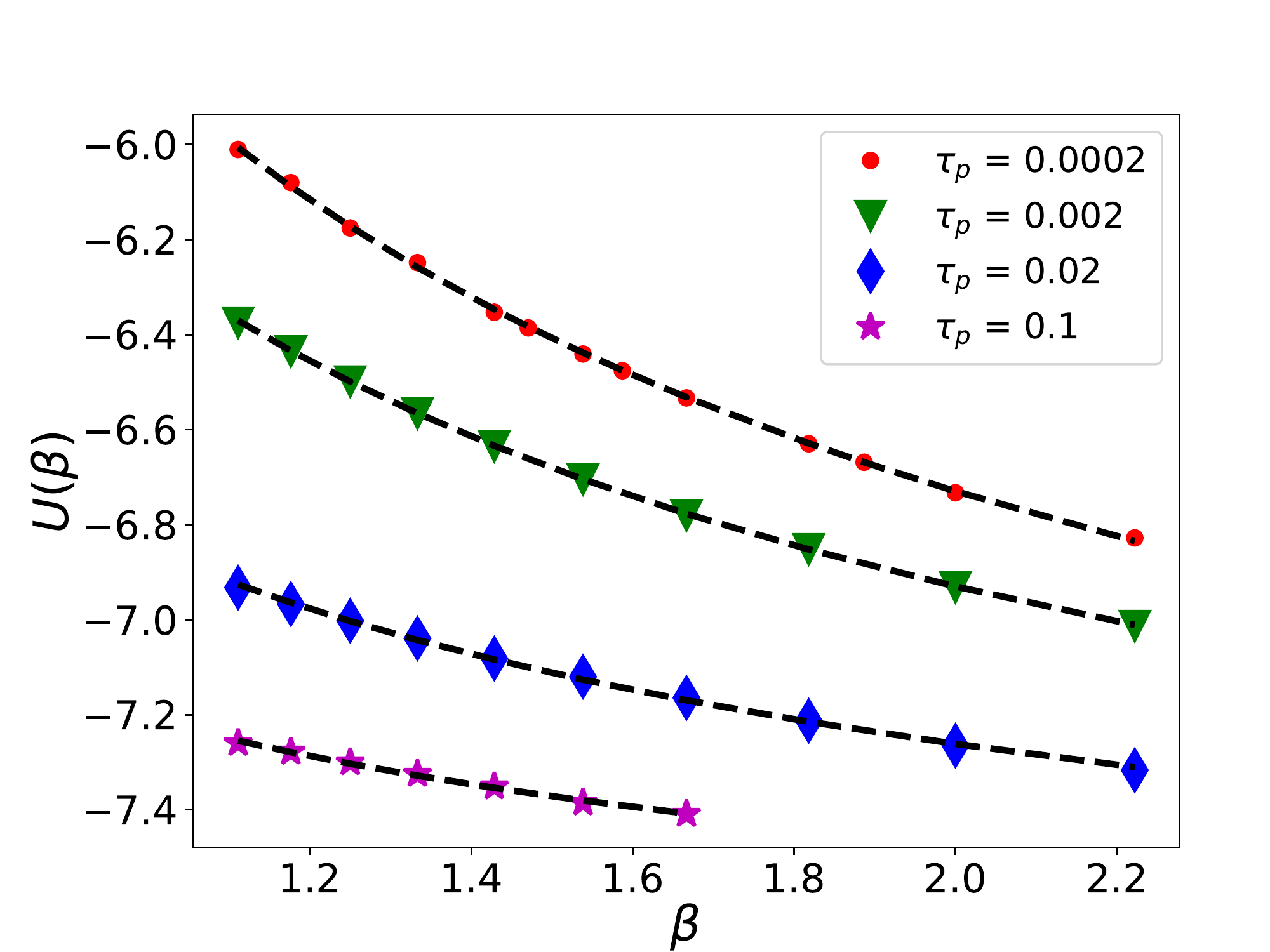}
      \caption{Temperature dependence of average potential energy at various values of persistence time. Dashed lines indicate fitting to Eq. \ref{RT-fit} with parameters listed in table \ref{pot-fit-params}.}
 	\label{avg-pot-eng-fig}
 \end{figure}

\begin{figure}[ht!]
\includegraphics[scale=0.4]{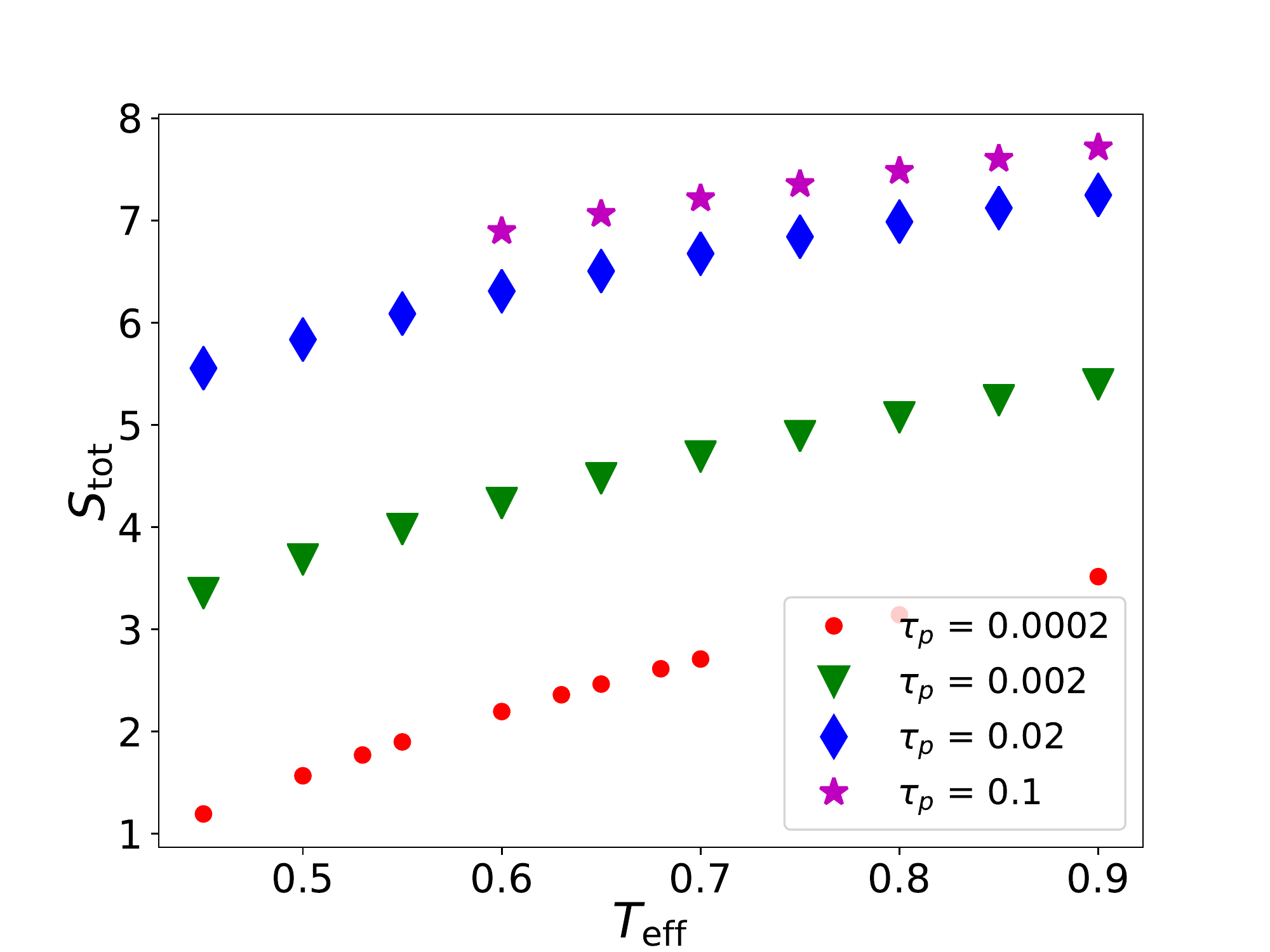}
\caption{Total entropy as a function of effective temperature at different values of $\tau_{\text{p}}$. It is increasing as a function of increasing $\tau_{\text{p}}$ at any fixed effective temperature.}
\label{S-tot-fig}
\end{figure}

This section describes the procedure to compute the total entropy per particle, a sum of the ideal gas entropy $S_{\text{id}}$ and the excess entropy $S_{\text{exc}}$. The  excess entropy at a given state point can be calculated via thermodynamic integration starting from a reference state where the exact entropy is known \cite{sciortino1999inherent, coluzzi2000lennard}. We take the limit $T_{\text{eff}} \rightarrow \infty$ as this reference state which is common practice for such calculations. The total entropy for any soft interaction such as the Lennard-Jones potential then reads as,
\begin{equation}
S_{\text{tot}} = S_{\text{id}} + \beta U(\beta) - \int_0^{\beta}\text{d}\beta' U(\beta')
\label{Stot-eq}
\end{equation}
where $U(\beta)$ is potential energy averaged over the ensemble at some inverse effective temperature $\beta = 1/T_{\text{eff}}$. The ideal gas entropy per particle can be evaluated as
\begin{equation}
S_{\text{id}} = \frac{d+2}{2} - \ln\rho - \ln\Lambda^d + S_{\text{mix}}
\end{equation}
where, $d$ is the spatial dimension, $\Lambda$ is the thermal wavelength and $S_{\text{mix}} = -\sum_{m=1}^{M}X_{m}\ln X_{m}$ is the mixing entropy per particle. Here $X_m$ are the concentration of $m^\text{th}$ species. The integral in the Eq. \ref{Stot-eq} is then evaluated by fitting the averaged potential energy with the Rosenfeld-Tarazona polynomial \cite{ingebrigtsen2013communication, das2022crossover}
\begin{equation}
  U(\beta) = a + b\beta^{-k}.
  \label{RT-fit}
\end{equation}
Figure \ref{avg-pot-eng-fig} shows this numerical fitting with the parameters $a$, $b$ and $k$ mentioned in table \ref{pot-fit-params}. Figure \ref{S-tot-fig} shows the variation of $S_{\text{tot}}$ as a function of $T_{\text{eff}}$ for various $\tau_{\text{p}}$ values and it is evident that at a fixed $T_{\text{eff}}$, increasing activity has the effect of increasing total entropy. Later in section \ref{section-CAG} we show how increasing the persistence time correlates positively with glassy behavior using the arguments of configurational entropy and energy barrier. In the following section we explore the PEL of this system and calculate $S_{\text{glass}}$ using the normal modes of vibrations.
 \begin{table}
 	\begin{tabular}{| c | c | c | c |}
 		\hline 
 		$\tau_{\text{p}}$ & $a$ & $b$ & $k$\\ 
 		\hline 
 		0.0002 & -8.27519 & 2.42532 & 0.648819 \\ 
 		\hline 
 		0.002 &  -8.3846& 2.10334 & 0.562469 \\ 
 		\hline 
 		0.02 & -9.44373 & 2.57571 & 0.240193\\
 		\hline
 		0.1 & -13.3553 & 6.13897 & 0.0627211\\
 		\hline
 	\end{tabular} 
	\caption{Fitting parameters $a$, $b$ and $k$ for different persistence times that are used in Eq. \ref{RT-fit}.}
 	\label{pot-fit-params}
 \end{table} 
 

\subsection{Glass Entropy from Inherent Structures}

\subsubsection{Inherent structures}
\begin{figure}[ht!]
\includegraphics[scale=0.4]{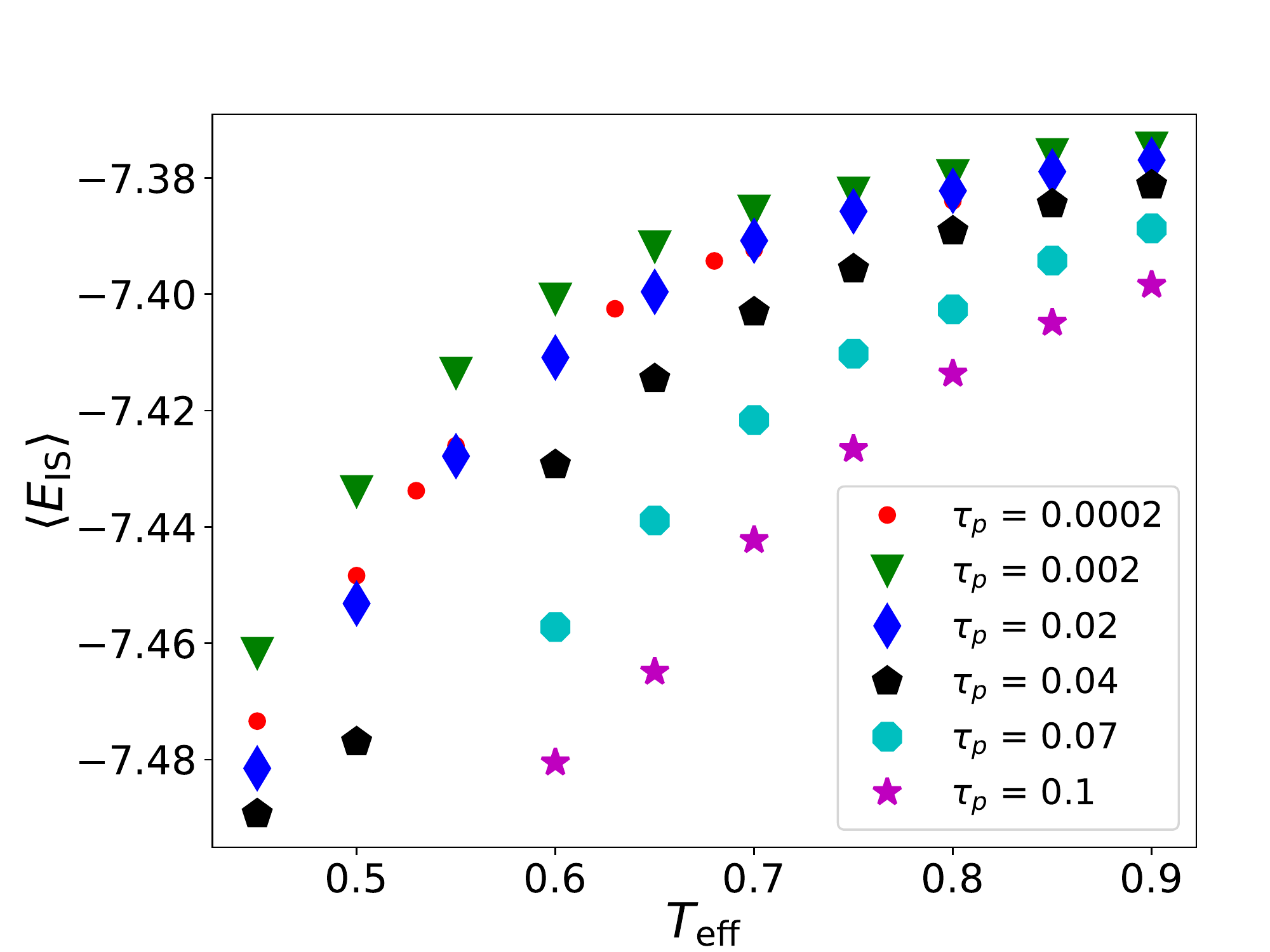}
\caption{Temperature dependence of the averaged inherent structure energies at various levels of activity. Here the data is averaged over 200 realizations and the standard deviation is of the order $\sim 10^{-2}$.}
\label{Avg-EIS}
\end{figure}

Inherent structure (IS) is a local minimum of the PEL that can provide important structural information of the glass forming liquids \cite{sastry2001relationship,sastry2002inherent,heuer2008exploring,sastry1999potential}. The multiplicity of such structures gives a direct measure of the configurational entropy. In our work, we used the steady state configurations at a given effective temperature and minimizing their energy using the conjugate gradient method \cite{shewchuk1994introduction,nocedal2006conjugate}. Figure \ref{Avg-EIS} shows the variation of averaged IS energy $\left< E_{IS} \right>$ as a function of the effective temperature plotted at various persistence times. It can be observed that at a particular $T_{\text{eff}}$,  $\left< E_{IS} \right>$ becomes negatively larger at large $\tau_{\text{p}}$ \textemdash implying the system samples deeper minima in the PEL.
The harmonic contribution to glass entropy $S_{\text{glass}}^{\text{harm}}$ is calculated by making a harmonic approximation to the potential energy close to the IS. This is explained further in the following section.

\subsubsection{Harmonic contribution to the glass entropy}

\begin{figure}[ht!]
\includegraphics[scale=0.4]{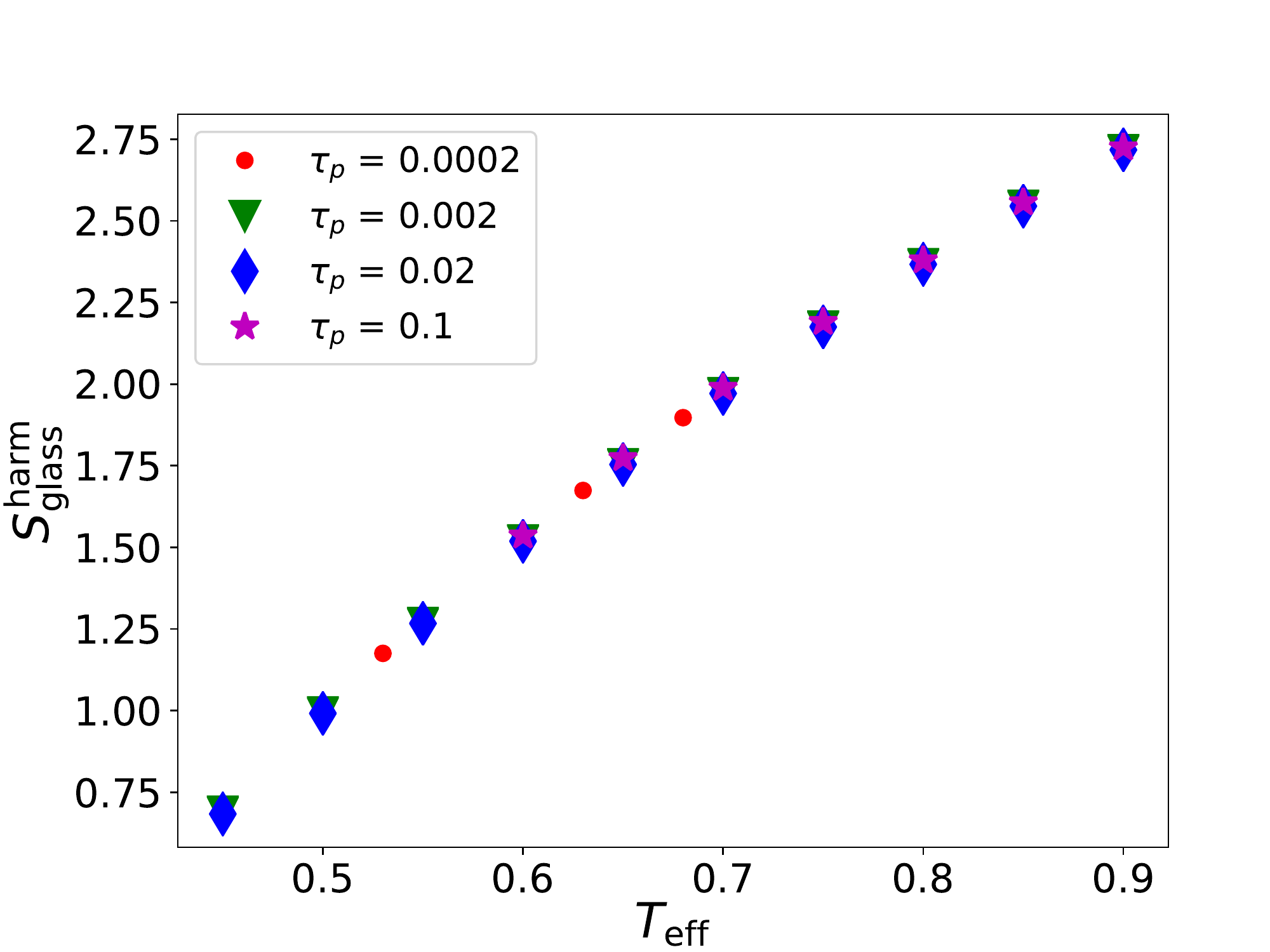}
\caption{Harmonic component of glass entropy as a function of $T_{\text{eff}}$. It is sensitive only to the effective temperature as can expected from Eq. \ref{S-glass-harm-eqn}.}
\label{S-harm-fig}
\end{figure}

We start with a typical steady state configuration and quench it to the nearest inherent structure via the conjugate gradient procedure \cite{wright1999numerical}. We then calculate the Hessian of these minimized states, which is a matrix that gives the second derivative of the potential energy \cite{karmakar2010athermal, das2022crossover}
\begin{equation}
  \mathcal{H}^{ij}_{\chi\zeta} = \frac{\partial^2 U}{\partial r^i_\chi r^j_\zeta}
\end{equation}
Clearly this is always symmetric and also positive semi-definite when computed for the minimized states. Its eigenvalues can therefore be only positive or zero, the latter originating from the symmetries of the system. We diagonalize the Hessian matrix using the LAPACK package \cite{anderson1999lapack} to obtain the normal modes of vibrations. This allows us to compute $S_{\text{glass}}^{\text{harm}}$ in terms of the normal modes as follows \cite{Sciortino_2005},
\begin{equation}
  S_{\text{glass}}^{\text{harm}} = \left< \sum_{n = 1}^{Nd} \{1-\beta \hbar \omega_n \}\right>_{\text{IS}}
  \label{S-glass-harm-eqn}
\end{equation}
with $\omega_n$ are the normal frequencies of vibrations. In figure \ref{S-harm-fig}, we plot this harmonic contribution as a function of effective temperature at various persistence times. As can be expected, the harmonic contribution is sensitive only to the variations in effective temperature but not the persistence time. In passive glass formers, this harmonic contribution usually plays the dominant role and the anharmonic contribution is almost always neglected in comparison \cite{berthier2019configurational}. We find this to be in stark contrast to active systems where the anharmonic contribution can become significant with activity and must be therefore included for the correct estimation of configurational entropy. This is discussed next. 

\subsubsection{Anharmonic contribution to the glass entropy}
\begin{figure}[ht!]
\includegraphics[scale=0.4]{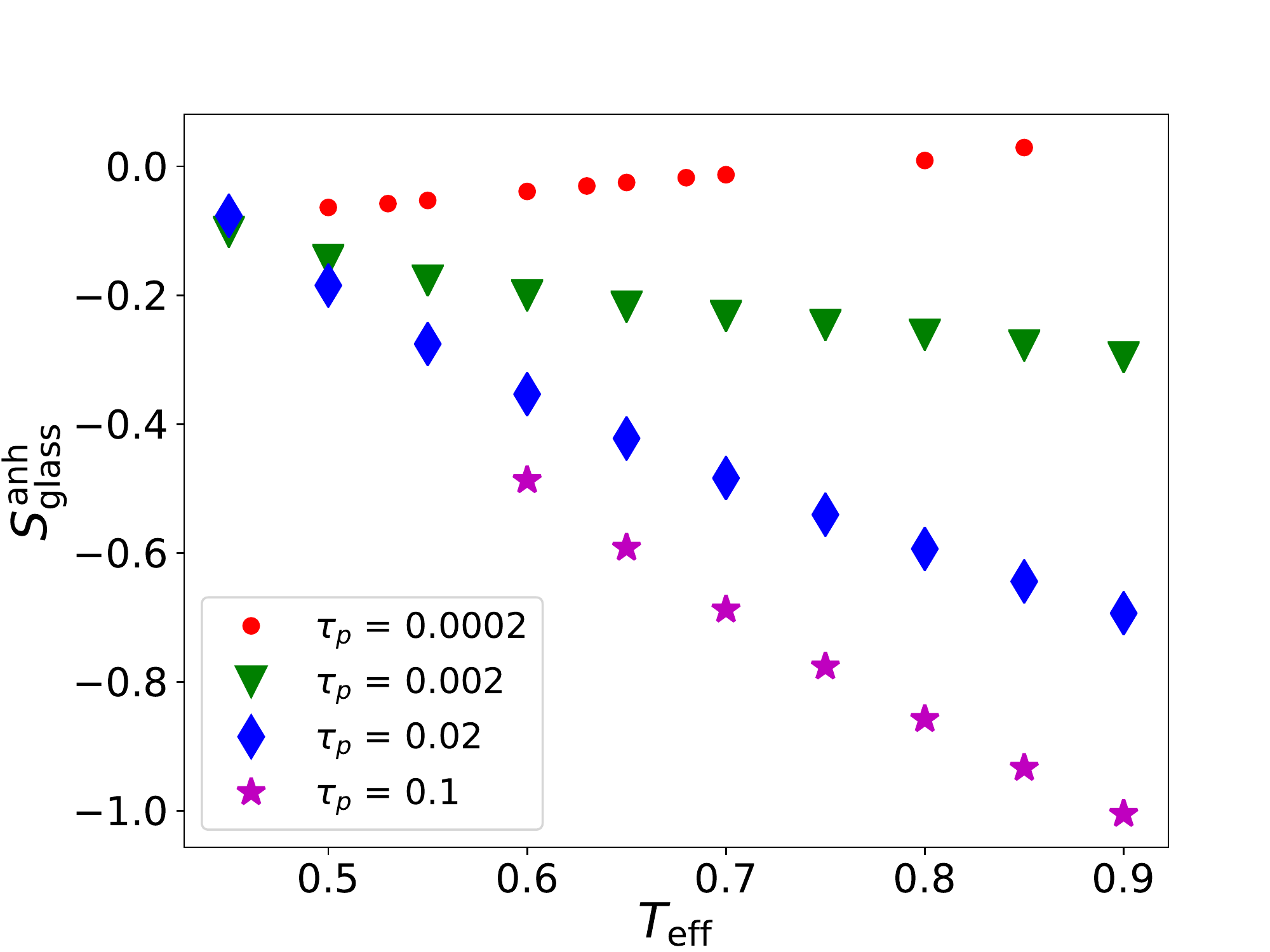}
\caption{Anharmonic contribution to the glass entropy as a function of $T_{\text{eff}}$. For a better representation we have shifted the curves for $\tau_{\text{p}} = 0.002, 0.02\; \text{and}\; 0.1$ along the vertical axis by 1.2, 3.2 and 3.9 respectively.}
\label{S-anharm-fig}
\end{figure}

 In order to calculate the anharmonic glass entropy $S_{\text{\text{glass}}}^{\text{\text{anh}}}$ \cite{Sciortino_2005}, we made use of the anharmonic energy $E_{\text{anh}}$ which is the difference between the total energy and the harmonic energy. The anharmonic contribution $S_{\text{glass}}^{\text{anh}}$ can be calculated using the relation,
\begin{equation}
S_{\text{glass}}^{\text{anh}} = \int_{0}^{T_{\text{eff}}}\frac{\text{d}T'_{\text{eff}}}{T'_{\text{eff}}}\frac{\partial E_{\text{anh}}(T'_{\text{eff}})}{\partial T'_{\text{eff}}}
\label{S-glass-anh-eq}
\end{equation}
where, the integral in the equation \ref{S-glass-anh-eq} is solved by fitting $E_{\text{anh}}(T_{\text{eff}})$ with a polynomial of the form $E_{\text{anh}}(T_{\text{eff}}) = \sum_{m \geq 2}a_mT_{\text{eff}}^m$. This results in the simplified form of the above equation given by,
\begin{equation}
S^{\text{anh}}_{\text{glass}} = \sum_{m \geq 2} \frac{m}{m-1} a_m T_{\text{eff}}^{m-1}
\end{equation} 
here, the lower bound for the summation is chosen in such a way that, its heat capacity should go to zero at $T_{\text{eff}} = 0$. Figure \ref{S-anharm-fig} shows the temperature variation of the anharmonic entropy for various values of persistence time. It is evident that the anharmonic contribution becomes negatively larger with increasing  persistence time at any given effective temperature. In the following section we compute the total glass entropy.

\subsubsection{Total glass entropy}
\begin{figure}[ht]
\includegraphics[scale=0.4]{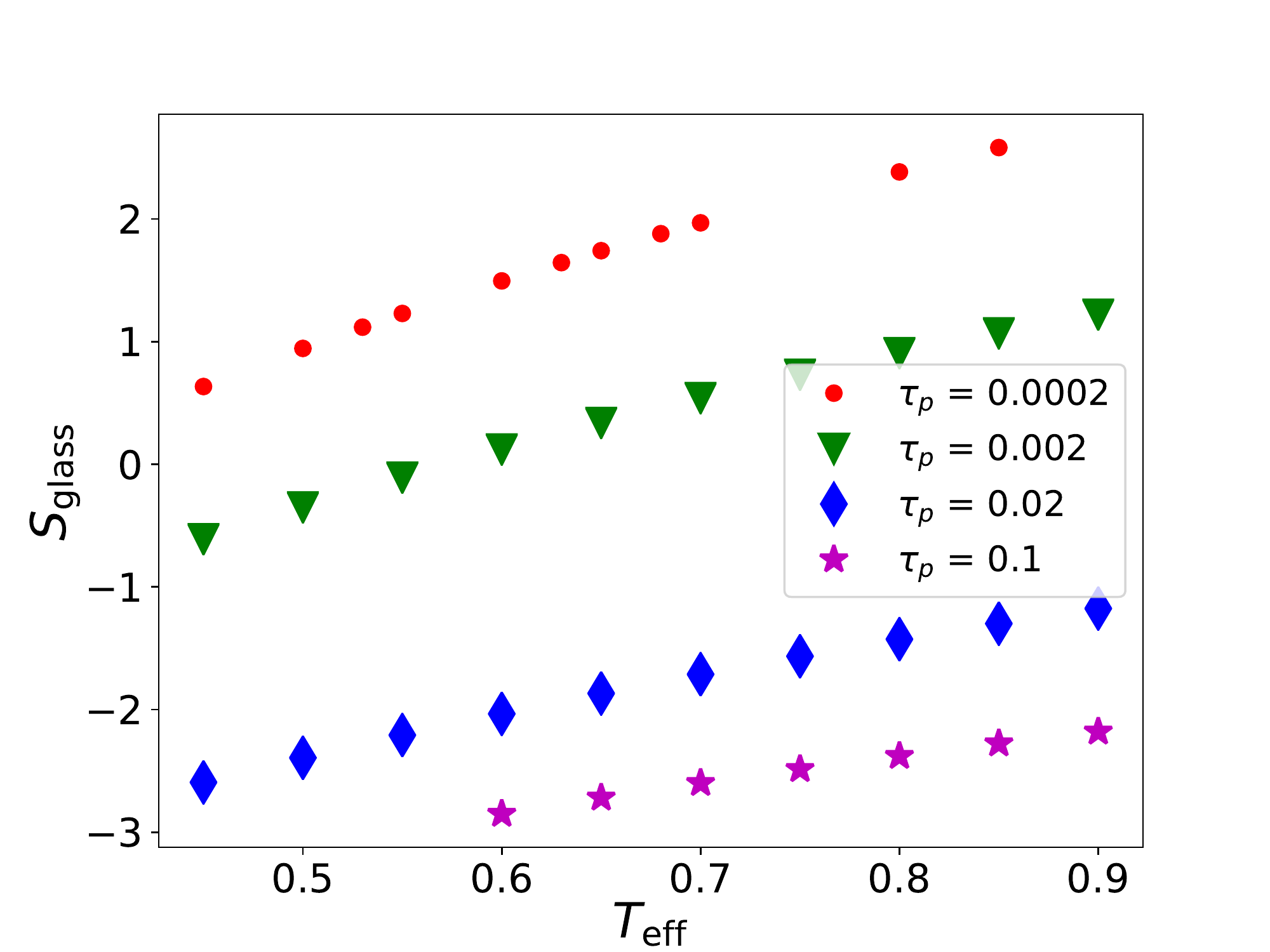}
\caption{Total glass entropy as a function of effective temperature at various levels of activity. Variation of glass entropy as a function of self-propulsion is entirely due to the anharmonic contributions discussed in the previous section.}
\label{S-glass}
\end{figure}

Having calculated the harmonic and anharmonic contributions to the glass entropy, we are now in a position to calculate the $S_{\text{glass}}$ which is a sum of both the contributions \cite{angelani2007configurational, sastry2000evaluation}.
\begin{equation}
S_{\text{glass}} = S_{\text{glass}}^{\text{harm}} + S_{\text{glass}}^{\text{anh}}
\end{equation}
Figure \ref{S-glass} shows the behavior of the total glass entropy as a function of effective temperature for different persistence times. At any fixed effective temperature, the variation in entropy with persistence time is entirely due to the anharmonic contribution  $S_{\text{glass}}^{\text{anh}}$ that was computed earlier. This asserts the importance and non-trivial nature of anharmonic effects in active systems that was stressed earlier in this paper. The results discussed here will enable us to compute the configurational entropy that can be used to describe the relaxation time of our low temperature active liquid. This is the subject matter of the next section.

\subsection{Generalized Adam-Gibbs relation in the self-propelled system}
\label{section-CAG}	

\begin{figure}
	\includegraphics[scale=0.4]{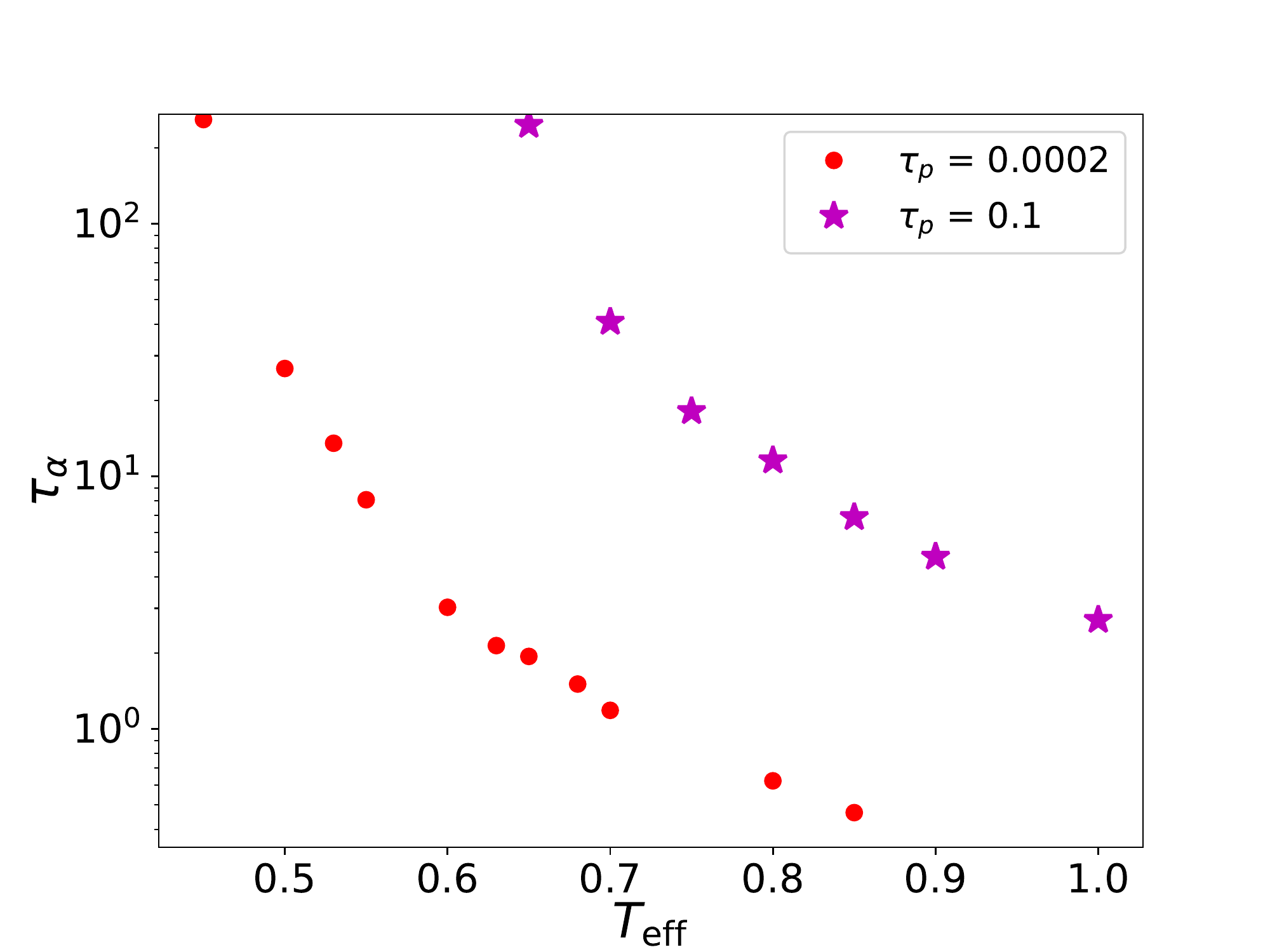}
	\caption{Temperature variation of relaxation time at two different levels of activity. Note that $\tau_{\text{p}} = 0.002$ corresponds to the Brownian limit. It is evident that any fixed $T_{\text{eff}}$, relaxation time increases with $\tau_{\text{p}}$.}
	\label{tau-alpha-T}
\end{figure}

\begin{figure}[ht!]
	\includegraphics[scale=0.4]{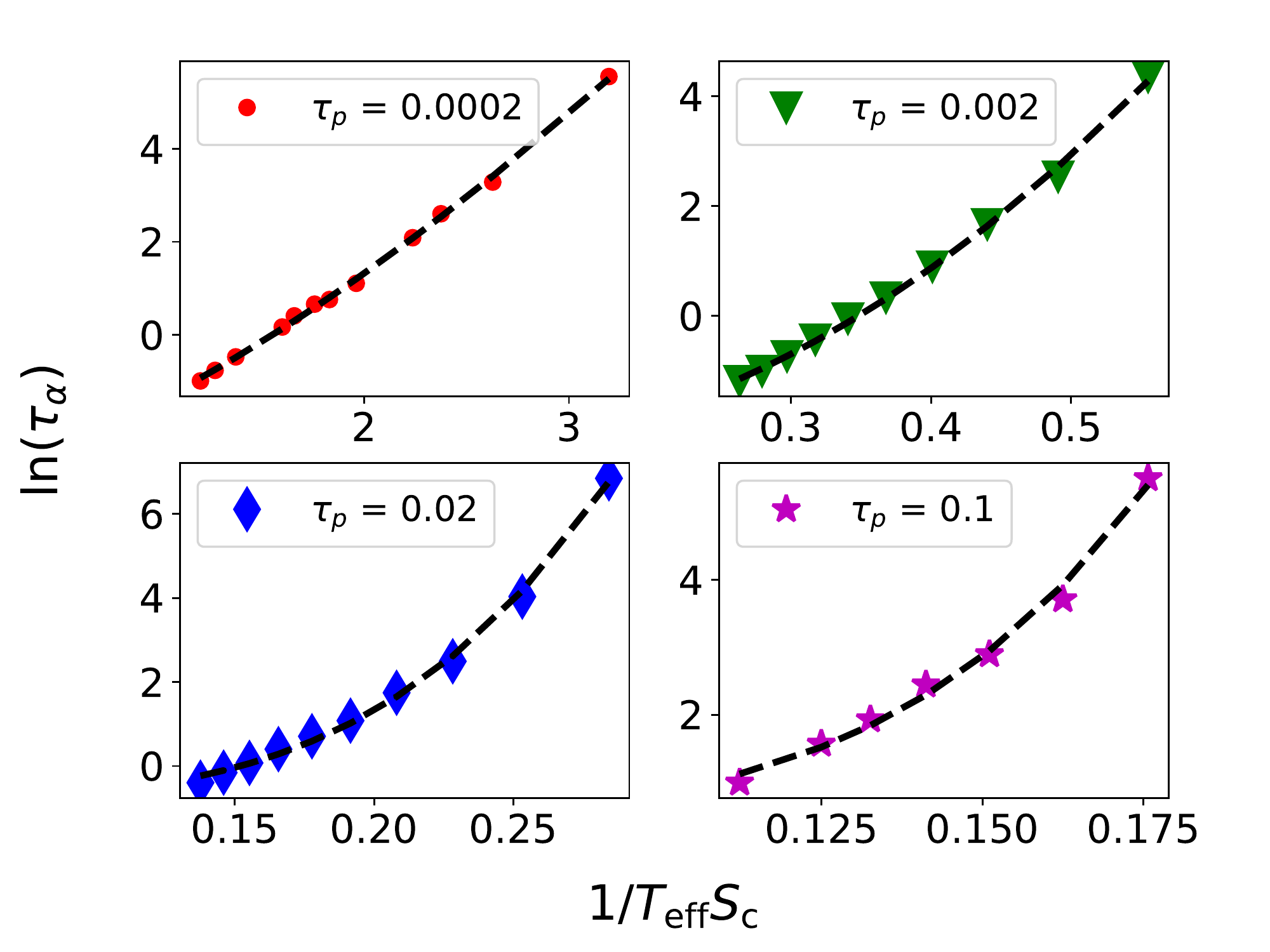}
	\caption{Verification of generalized Adam-Gibbs relation for the various values of persistence time. Dashed line represents the scaling relation given in the eq. \ref{AG-eq}. The extracted $\Delta E$ and $\delta$ are given in  table \ref{Ebarrier-table}.}
	\label{AG-rel-fig}
\end{figure}

 RFOT offers a rationalized view of AG theory based on the activated dynamics. This generalization describes the glass transition via a thermodynamic route that connects the relaxation time $\tau_{\alpha}$ to the configurational entropy through the following relation \cite{bouchaud2004adam, ozawa2019does},
\begin{equation}
        \tau_{\alpha} \sim \text{exp}\left[\left(\frac{\Delta E}{T_{\text{eff}}S_{\text{c}}}\right)^{\delta}\right]
        \label{AG-eq}
\end{equation}
where $\Delta E$ is the energy barrier and $\delta$ is a non universal constant, both varying with the persistence time. This deviation from the actual AG theory ($\delta$ = 1) was observed in previous studies \cite{sengupta2012adam}. The calculation of $S_{\text{c}}$ is done using equation \ref{S-conf-eq} that was discussed earlier and is shown in the figure \ref{S-conf-fig} below. At a fixed $\tau_{\text{p}}$, reducing $T_{\text{eff}}$ leads to a reduction in the number of IS resulting in slower relaxation of the system (see figure \ref{tau-alpha-T}) on the other hand, $S_{\text{c}}$ grows with $\tau_{\text{p}}$ at a fixed $T_{\text{eff}}$.
The latter has the effect of enhancing energy barrier height as a function of $\tau_{\text{p}}$ \textemdash an effect that is consistent with the growth of relaxation time with increasing persistence time at any fixed effective temperature (see figure \ref{tau-alpha-T}). Table \ref{Ebarrier-table} shows the behavior of fitting parameters obtained from the generalized Adam-Gibbs relation obeyed by our data for different persistence times as shown in the figure \ref{AG-rel-fig}. 
 
\begin{figure}[ht!]
	\includegraphics[scale=0.4]{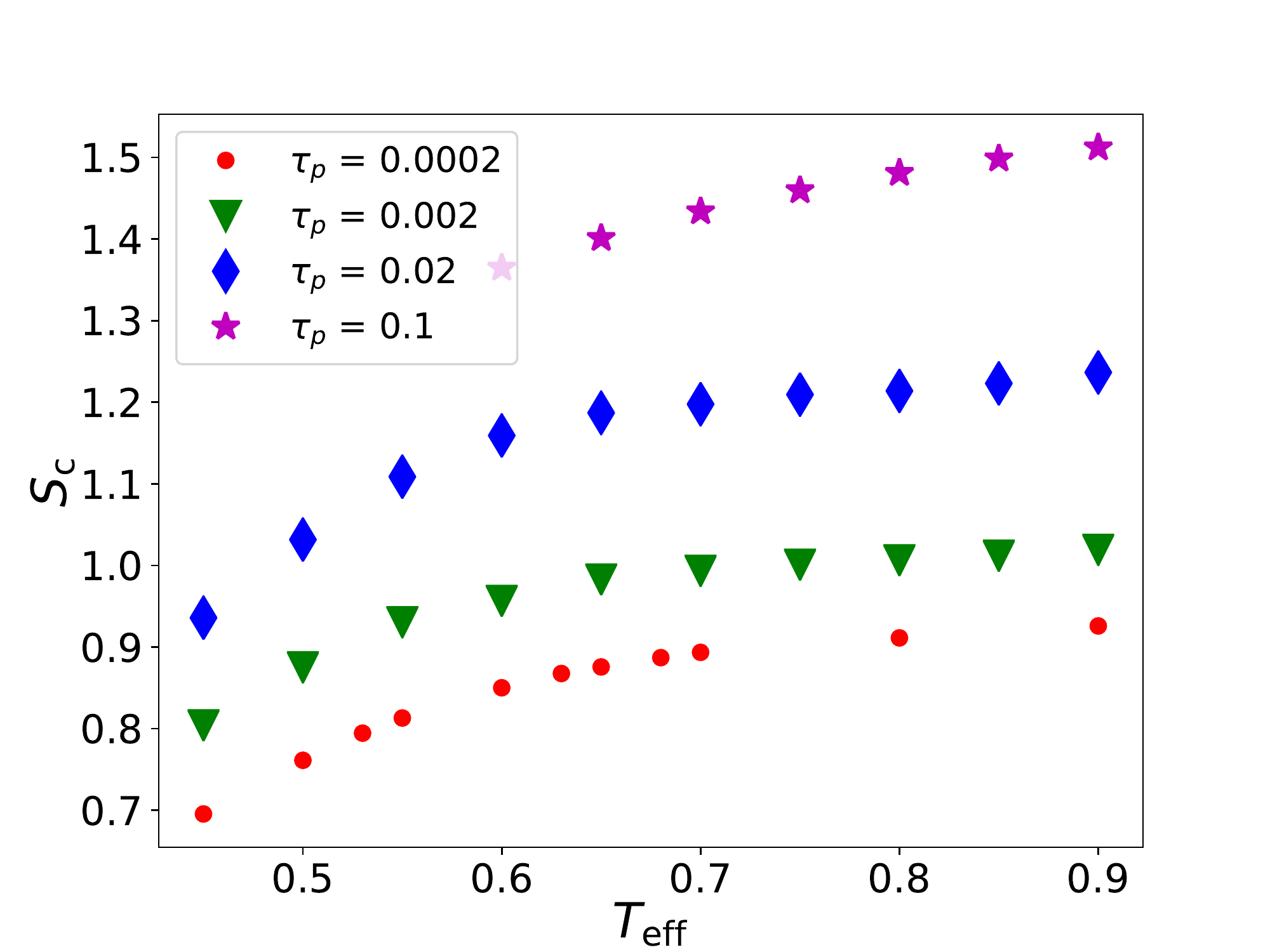}
	\caption{Configurational entropy as a function of effective temperature at different values of $\tau_{\text{p}}$. At a fixed $\tau_{\text{p}}$, reducing $T_{\text{eff}}$ reduces $S_{\text{c}}$. On the other hand, $S_{\text{c}}$ increases as a function of $\tau_{\text{p}}$ at a fixed $T_{\text{eff}}$.   For a better representation we have shifted the curves for $\tau_{\text{p}} = 0.002, 0.02 \; \text{and}\; 0.1$ along the vertical axis by -3.20, -6.80 and -8.00 respectively.}
	\label{S-conf-fig}
\end{figure}
\begin{table}[ht!]
	\begin{tabular}{|c|c|c|}
		\hline
	$\tau_{\text{p}}$ & $\Delta E$ & $\delta$ \\
		\hline
	0.0002	&  1.280 & 1.503 \\
		\hline
	0.002	&  4.520 & 2.095 \\
		\hline
	0.02	&  6.129 & 3.636 \\
		\hline
	0.1	&  8.041 & 4.605 \\
		\hline
	\end{tabular}
\caption{The values of energy barrier and the degree of deviation from actual AG relation ($\delta$) at various persistence times obtained as fitting parameters from equation \ref{AG-eq}.}
\label{Ebarrier-table}
\end{table}
The reader should note that the results reported here are obtained without recoursing to a first order approximation of configurational entropy \cite{nandi2018random}.
Further we went ahead and compute a static length scale $\xi_{\text{s}}$ by a random pinning of particles and deduced an RFOT like scaling law between $S_{\text{c}}$ and $\xi_{\text{s}}$. This is discussed below.

\section{Static length scale and the configurational entropy}
\label{LenScale-section}
\begin{figure}[!]
	\centering
	\begin{minipage}[b]{0.5\textwidth}
		\centering
		\includegraphics[scale=0.4]{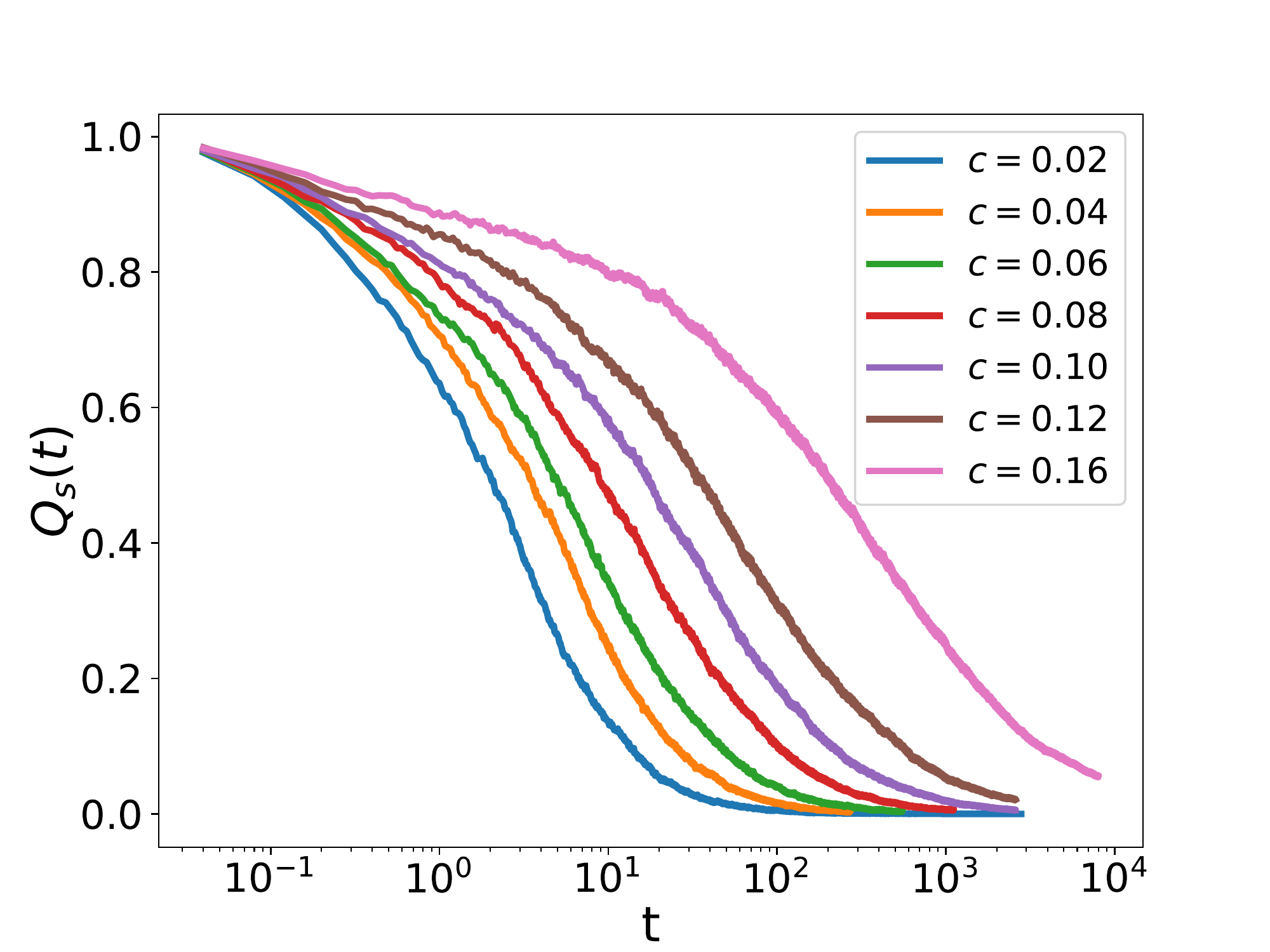}
	\end{minipage}
	\hfill
	\begin{minipage}[b]{0.5\textwidth}
		\centering
		\includegraphics[scale=0.4]{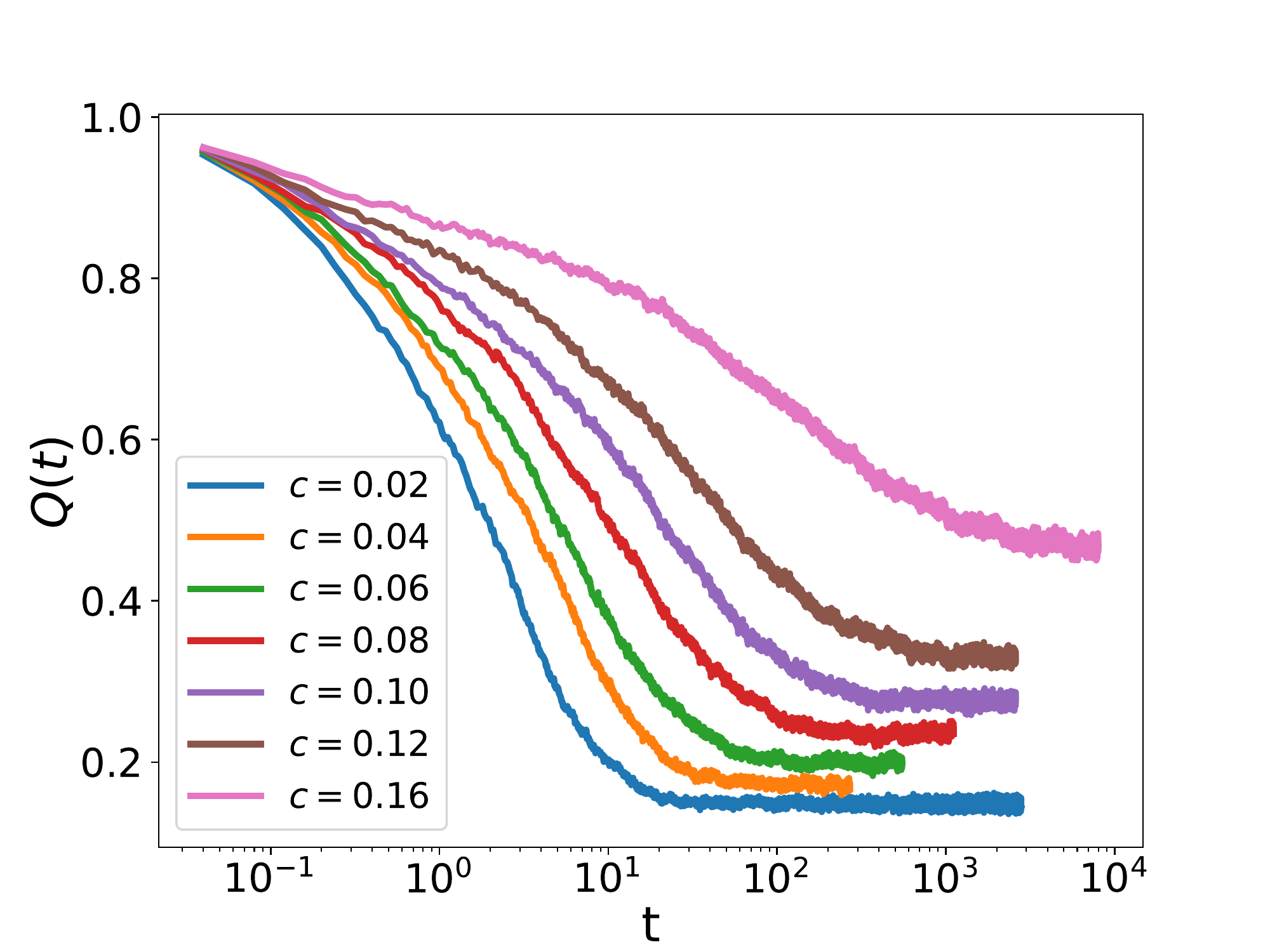}
	\end{minipage}
	\caption{The top and the bottom figures show the time decay of the self part and the total part of the overlap function respectively as a function of time for various pinning concentration at $T_{\text{eff}} = 0.65$ and $\tau_{\text{p}} = 0.002$. The asymptotic value of the total overlap function $Q_{c}(\infty)$ is measured when it saturates to a non-zero value.}
	\label{OPF-fig}
\end{figure}

The RFOT for the passive system suggests that liquid is composed of metastable regions called ``mosaics'' with a characteristic length scale $\xi_{\text{s}}$ \cite{kirkpatrick1987connections,kirkpatrick1989scaling,lubchenko2007theory,biroli2012random,kirkpatrick2015colloquium}. The reorganization of these mosaic regions depends on the energy barrier which scales as $\xi_{\text{s}}^{\psi}$ \cite{starr2013relationship, karmakar2009growing}. This directly provides a scaling relation between relaxation time and the length scale as follows,

\begin{figure}[!]
	\centering
	\begin{minipage}[b]{0.5\textwidth}
		\centering
		\includegraphics[scale=0.2]{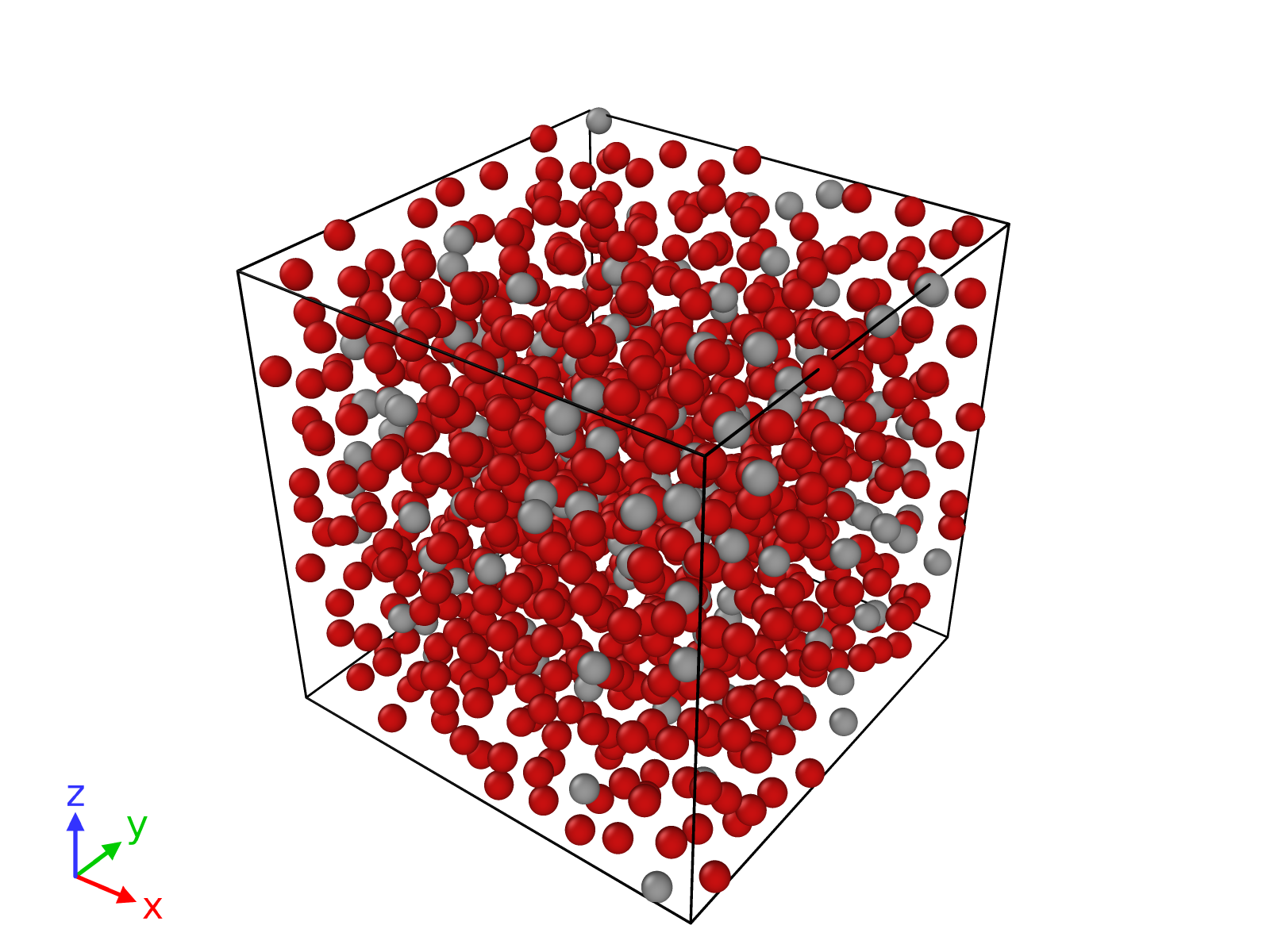}
	\end{minipage}
	\hfill
	\begin{minipage}[b]{0.5\textwidth}
		\centering
		\includegraphics[scale=0.2]{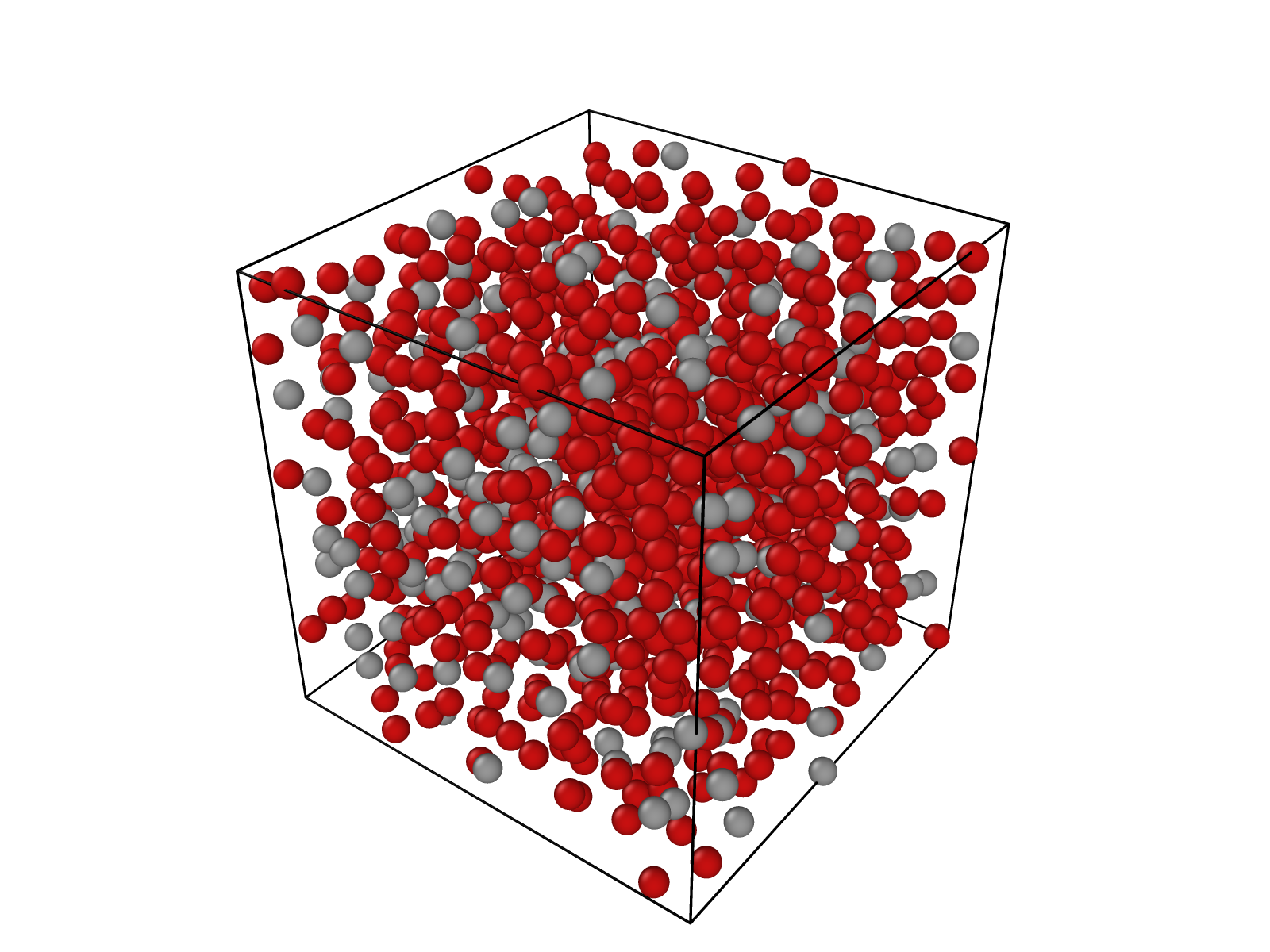}
	\end{minipage}
	\hfill
	\begin{minipage}[b]{0.5\textwidth}
		\centering
		\includegraphics[scale=0.2]{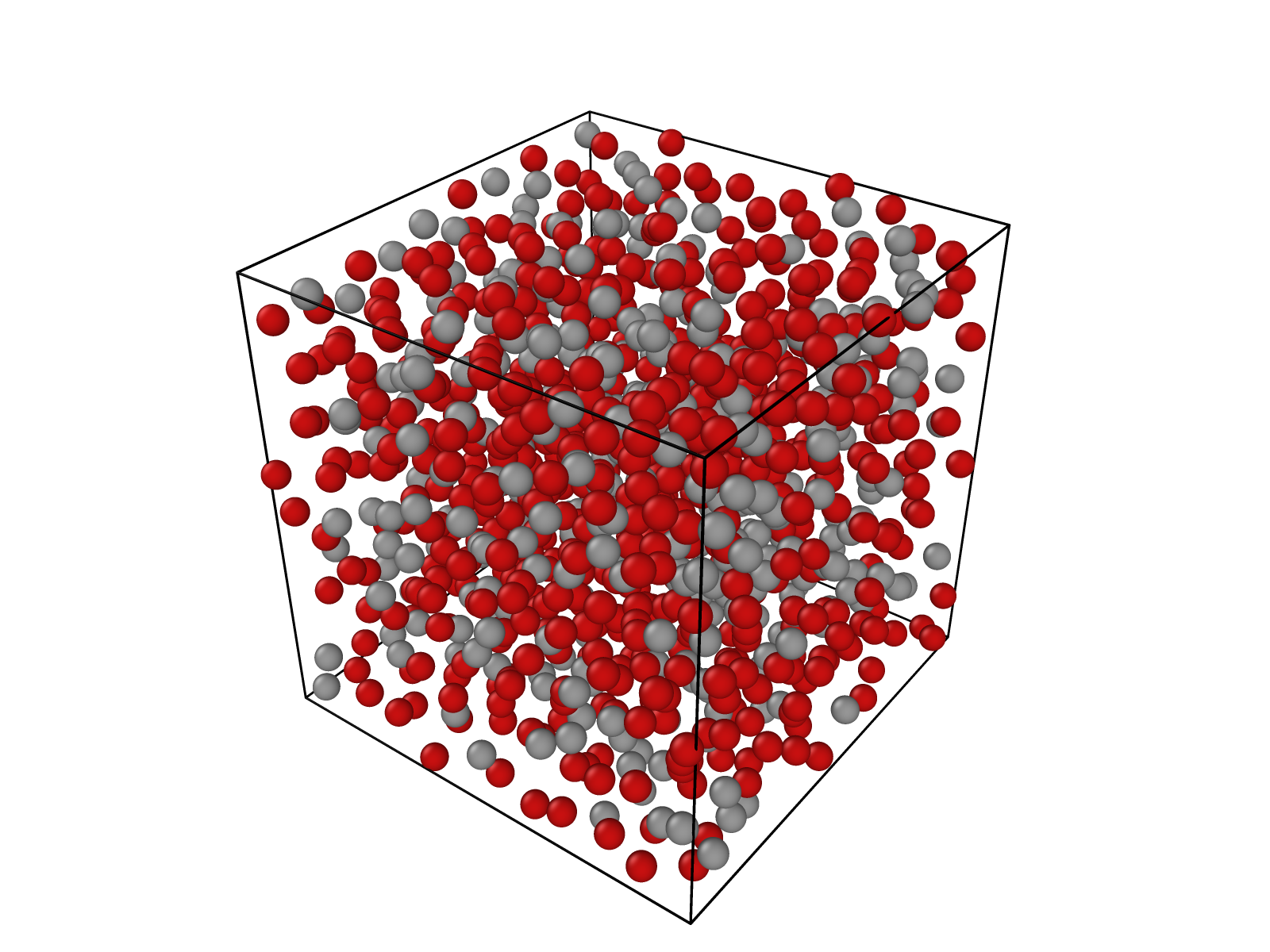}
	\end{minipage}
	\caption{Random pinning of the equilibrated configuration at an effective temperature $T_{\text{eff}} = 1.0$ and $\tau_{\text{p}} = 0.0002$ with a pinning concentrations of $c$ = 0.16, 0.25 and 0.35 (from top to bottom). Here, the Grey particles are pinned while the red particles are unpinned.}
	\label{RP-fig}
\end{figure}

\begin{equation}
	\tau_{\alpha} \sim \text{exp}\left(\frac{A\; \xi_{s}^{\psi}}{T_{\text{eff}}}\right)
	\label{tau-alpha-xi-eq}
\end{equation}

where $A$ is a constant. To extract the PTS static length scale $\xi_{\text{s}}$ for self-propelled system we randomly choose particles in a steady state configuration and pin them such that they are distributed throughout the simulation region without any bias \cite{berthier2012static, chakrabarty2016understanding}. Figure \ref{RP-fig} shows such configurations with a pinning concentrations $c$ = 0.16, 0.25 and 0.35 (from top to bottom).  Using these pinned configurations as an initial state, we perform the dynamics for different $T_{\text{eff}}$ and $\tau_{\text{p}}$. All the runs are long enough to make sure that the self part of the overlap function (see eq. \ref{qt-self-eq}) decays to zero. The simulations are carried out for different pinning concentration. However, it should be noted that for each persistence time, at the low effective temperatures it is increasingly difficult to equilibrate the system as $c$ becomes larger. We calculate the overlap functions between the configurations $\textbf{r}_i(t)$ and $\textbf{r}_j(t)$ which facilitate the extraction of length scale as follows \cite{chakrabarty2016understanding, lavcevic2003spatially, nandi2022thermodynamics}.

\begin{equation}
	Q(t) = \frac{1}{N - N_p}\left[ \left< \sum_{i,j= 1}^{N-N_p} \omega(|\textbf{r}_i(t) - \textbf{r}_j(0)|)\right> \right]
	\label{qt-total-eq}
	\end{equation}
where $N_p$ is the number of pinned particles and the $\omega$ is a step function such that $\omega(x) = 1$ if $x < 0.30$ otherwise it is zero. Also, the self part of the overlap function reads as,
\begin{equation}
	Q_s(t) = \frac{1}{N - N_p}\left[ \left< \sum_{i = 1}^{N-N_p} \omega(|\textbf{r}_i(t) - \textbf{r}_i(0)|)\right> \right]
	\label{qt-self-eq}
\end{equation}
The figure \ref{OPF-fig} shows the behavior of total overlap function and its self part at an effective temperature $T_{\text{eff}} = 0.65$ and the persistence time $\tau_{\text{p}} = 0.002$. It is observed that, as the pinning concentration is increased, the relaxation time increases dramatically which makes the equilibration of the system a difficult task (see fig \ref{OPF-fig}). 

\begin{figure}[!]
	\centering
	\begin{minipage}[b]{0.5\textwidth}
		\centering
		\includegraphics[scale=0.4]{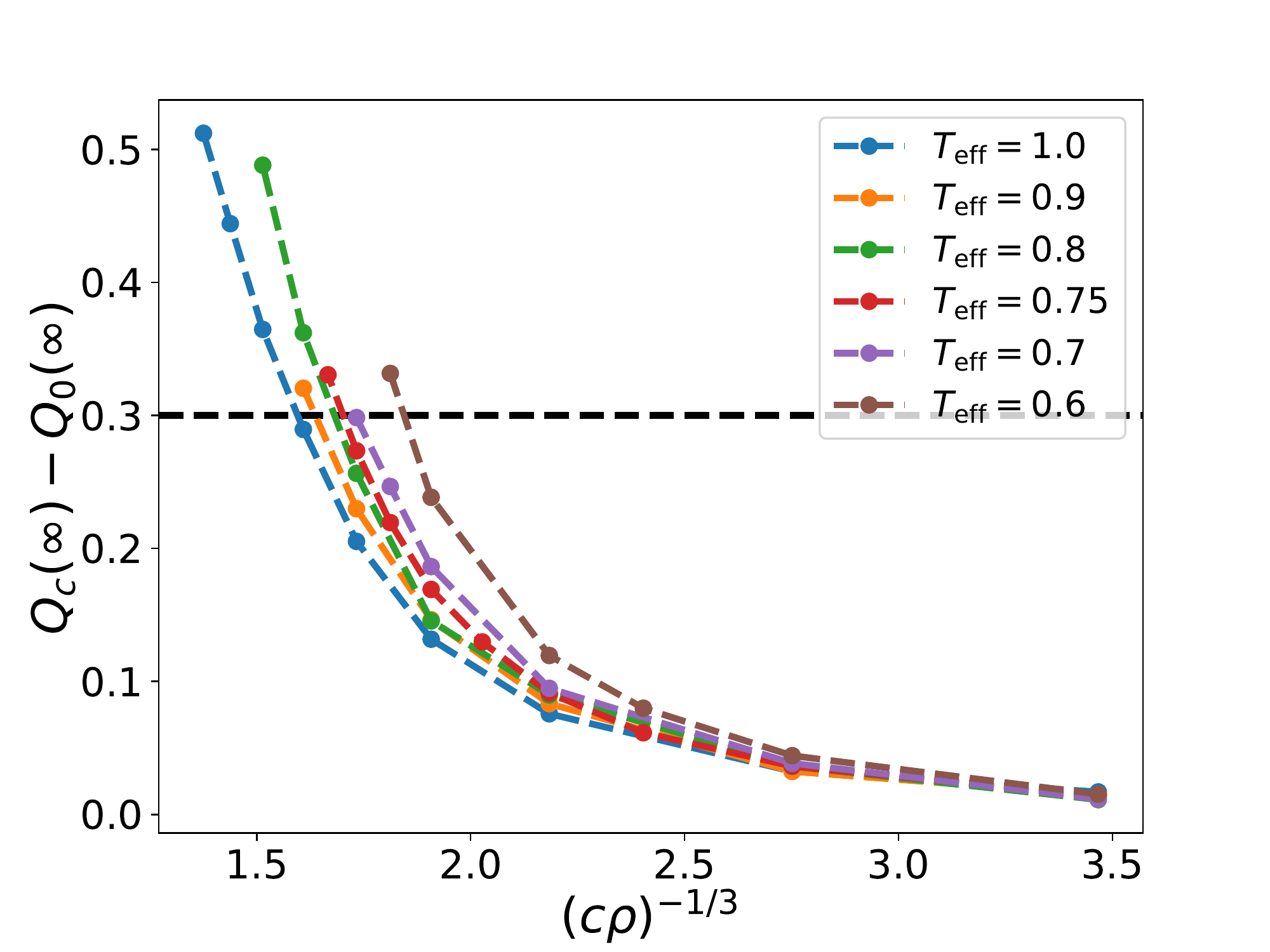}
	\end{minipage}
	\hfill
	\begin{minipage}[b]{0.5\textwidth}
		\centering
		\includegraphics[scale=0.4]{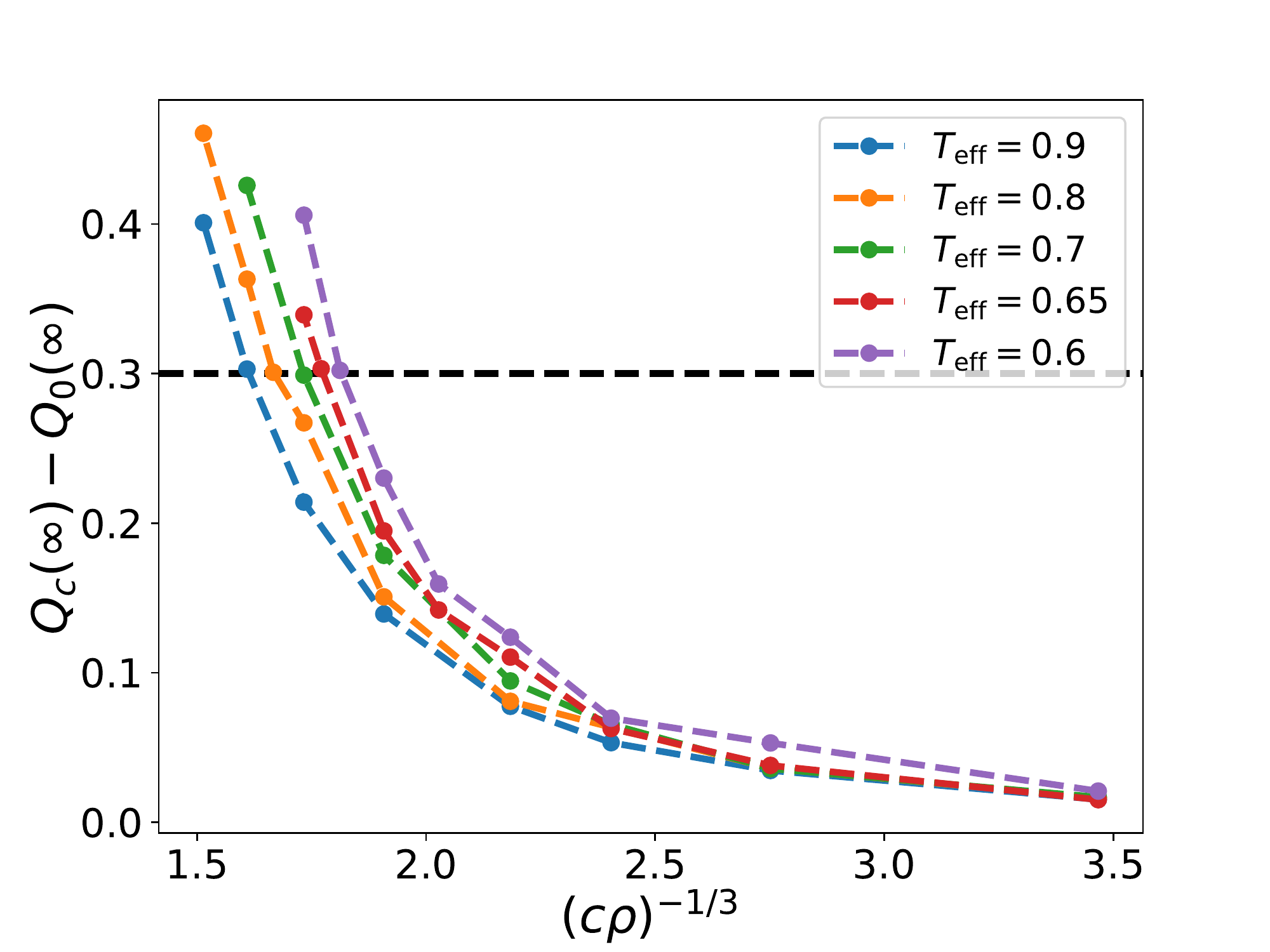}
	\end{minipage}
	\hfill
	\begin{minipage}[b]{0.5\textwidth}
		\centering
		\includegraphics[scale=0.4]{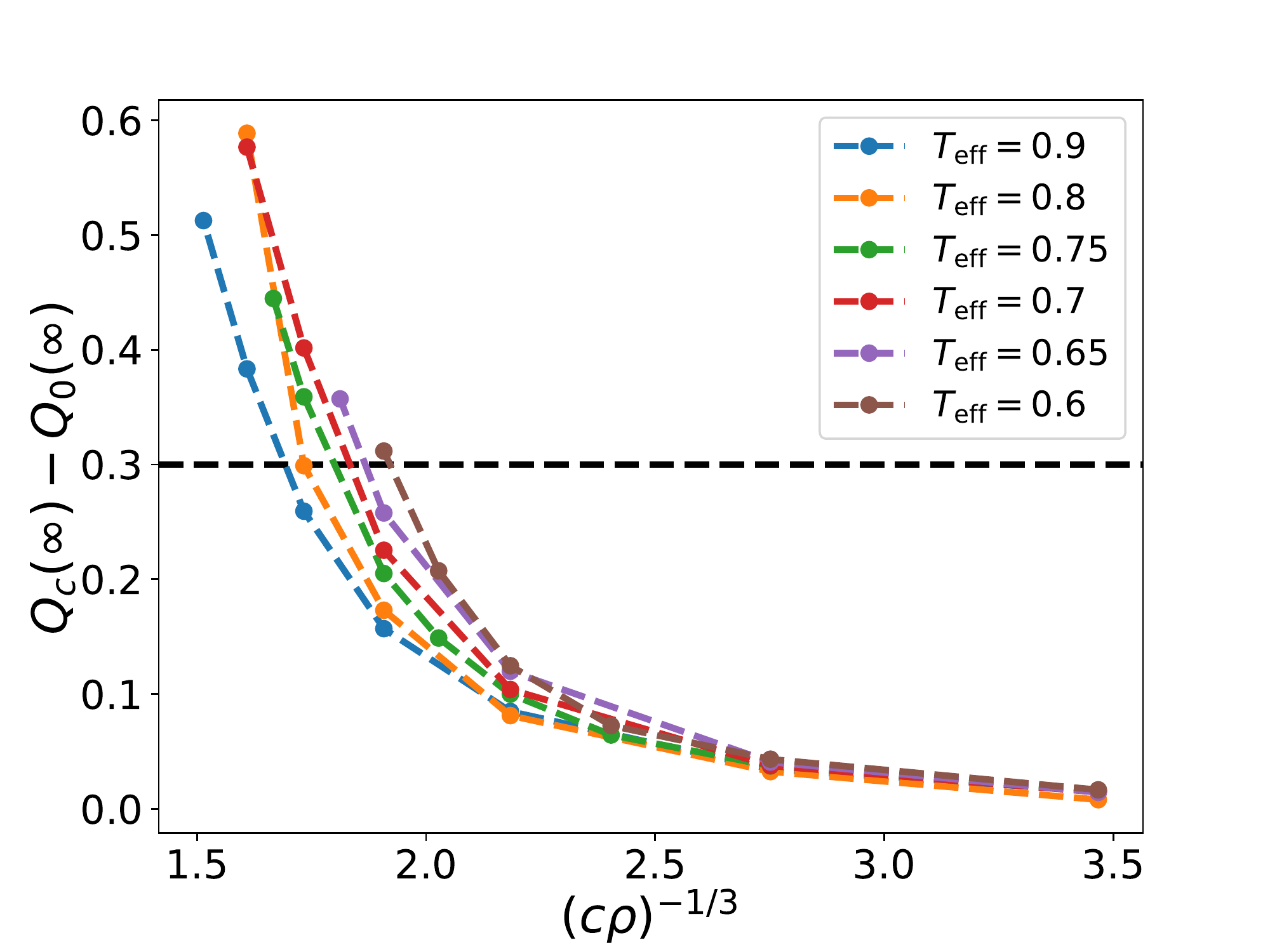}
	\end{minipage}
	\caption{The variation of difference between the asymptotic values of overlap function as a function of the average distance between the pinned particles. The shown plots are for $\tau_{\text{p}} = 0.0002, 0.002$ and $0.02$ from top to bottom. In each case the length scale $\xi_{\text{s}}$ is calculating value of $(c\rho)^{-1/3}$ corresponding to $Q_c(\infty) - Q_0(\infty)$ = 0.3 which is marked by the horizontal dotted line.}
	\label{OPF-asymp-diff-fig}
\end{figure}

\begin{figure}[!]
	\includegraphics[scale=0.4]{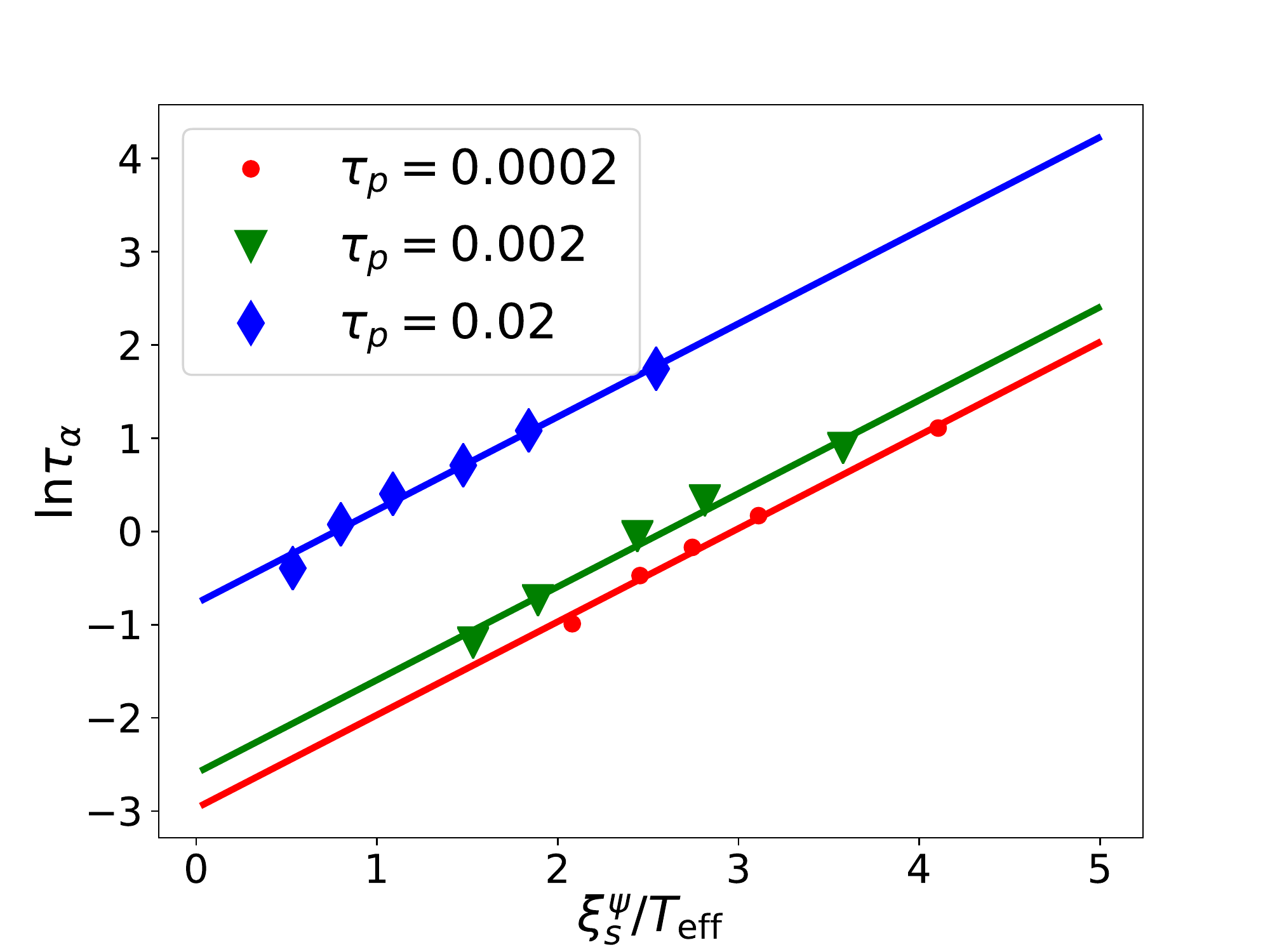}
	\caption{ The scaling law between the relaxation time $\tau_{\alpha}$ with the mosaic length scale $\xi_{\text{s}}$ for various values of $\tau_{\text{p}}$. The solid curves represents the scaling law given by \ref{tau-alpha-xi-eq}.}
\label{tau-xi-fig}	
\end{figure}

While the self part of the overlap function decays to zero, the total overlap function saturates at a non zero value in the asymptotic limit depending on pinning concentration. We denote this asymptotic value of the total overlap function as $Q_c(\infty)$, where the subscript $c$ represents the pinning concentration. In order to extract the length from random pinning, we construct a quantity $Q_c(\infty) - Q_0(\infty)$  with $Q_0(\infty)$ being the asymptotic value of overlap function with zero pinning concentration \cite{charbonneau2013decorrelation}. The variation of the above constructed quantity as a function of average distance between the pinned particles $(c\rho)^{-1/3}$ is shown in the figure \ref{OPF-asymp-diff-fig}. Now, the length scale $\xi_{\text{s}}$ is extracted as the value of $(c\rho)^{-1/3}$ corresponding to $Q_c(\infty) - Q_0(\infty)$ = 0.3 at different $T_{\text{eff}}$.
 Figure \ref{tau-xi-fig} shows the validity of the scaling relation (eq. \ref{tau-alpha-xi-eq}) for various persistence values. Values of $A$ and the exponent $\psi$ is given in the following table \ref{tau-xi-table}.

\begin{figure}
	\includegraphics[scale=0.4]{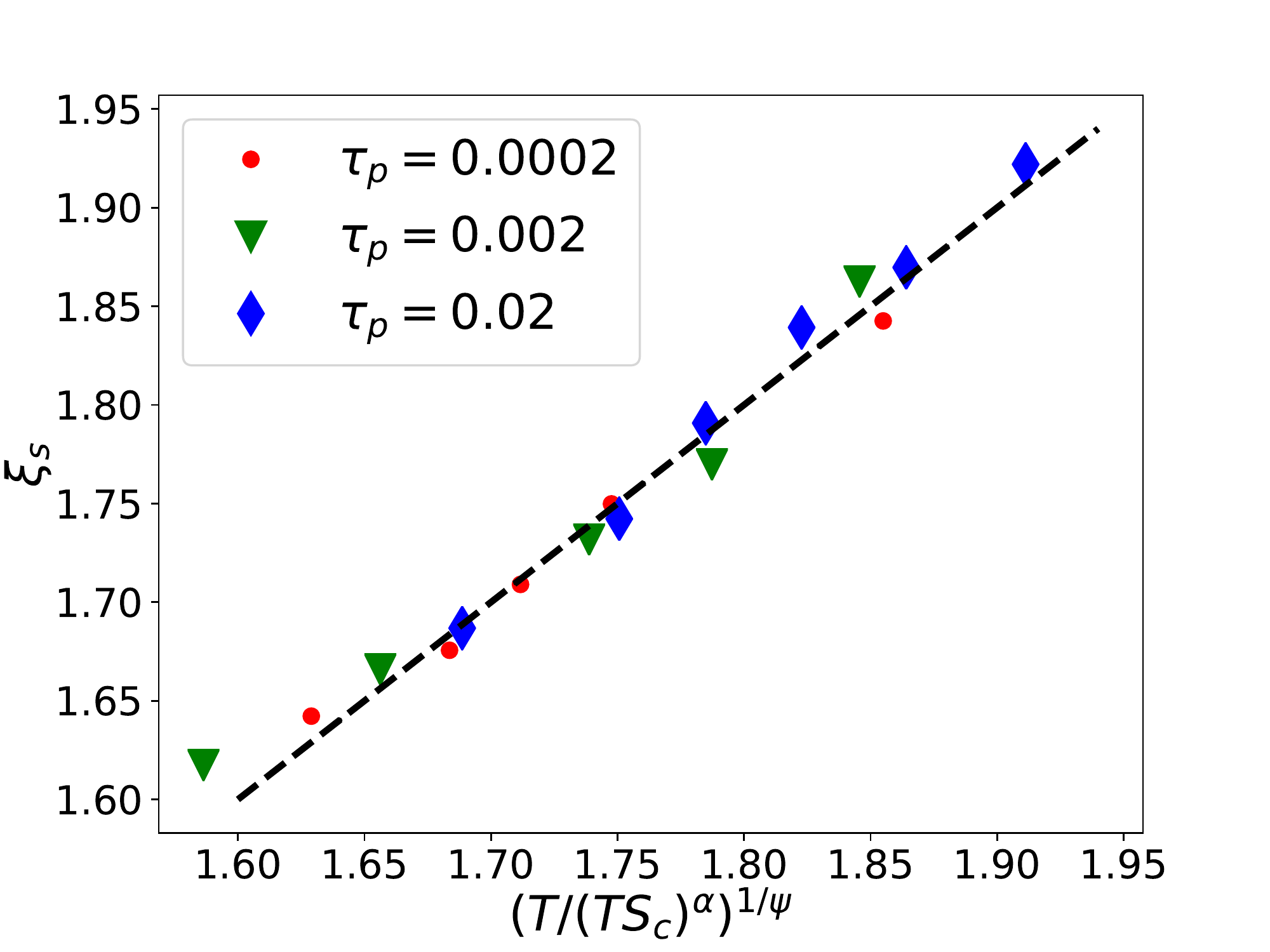}
	\caption{ Verification of the scaling law between $S_{\text{c}}$ and the length scale $\xi_{\text{s}}$ for various values of persistence time $\tau_{\text{p}}$. Data for different $\tau_{\text{p}}$ collapsed on the master curve (dashed line) which is governed by the equation deduced (eq. \ref{sc-xi-scaling-eq}).}
	\label{Sc-L-fig}
\end{figure}
\begin{table}[ht!]
	\begin{tabular}{| c | c | c |}
		\hline
		$\tau_{\text{p}}$ & $A$ & $\psi$ \\
		\hline
		0.0002	& 0.574 & 2.382 \\
		\hline
		0.002 	& 0.304 & 3.141 \\
		\hline
		0.02	& 0.007 & 8.859\\
		\hline
	\end{tabular}
	\caption{The values of fitting parameters $A$ and $\psi$ as a function of persistence time obtained from the equation \ref{tau-alpha-xi-eq}.}
	\label{tau-xi-table}
\end{table}

Equations \ref{AG-eq} and \ref{tau-alpha-xi-eq} enable us to deduce a scaling relation between $S_{\text{c}}$ and $\xi_{\text{s}}$ by eliminating $\tau_{\alpha}$ given by
\begin{equation}
	\xi_{\text{s}} = \left(\frac{\Delta E}{A} \right)^{1/\psi}\; \left(\frac{T_{\text{eff}}}{(T_{\text{eff}}S_{\text{c}})^{\delta}} \right)^{1/\psi}
	\label{sc-xi-scaling-eq}
\end{equation}
here, the effect of activity is to merely change the slope $\left(\frac{\Delta E}{A} \right)^{1/\psi}$. The validity of this scaling relation is presented in figure \ref{Sc-L-fig}, where the data is collapsed on to a master curve (dashed line) that is given by the equation \ref{sc-xi-scaling-eq}. Below we summarize our observations. 	

\section{Conclusion}
In this paper, we have provided an in detail calculation of the configurational entropy in the case of an athermal active glass former. Unlike passive liquids, the anharmonic contributions to configurational entropy are seen to be significant and also progressively increase with the degree of self-propulsion. The configurational entropy and relaxation time data for these self-propelled system satisfy the generalized for Adam-Gibbs relationship predicted by the random first order transition theory \textemdash an observation that has not been  reported so far in the literature. Our data shows the effect of lowering the effective temperature at a fixed $\tau_{\text{p}}$ is to reduce the number of available states resulting in slower relaxation of the system. On the other hand, increasing $\tau_{\text{p}}$ at a fixed $T_{\text{eff}}$ increases the energy barrier concomitant with activity enhancing the glassy behavior in AOUP system. We established an exponential scaling relation between the relaxation time and the PTS length scale which along with the generalized AG relation enabled  us to provide a direct relation between the PTS length scale $\xi_{\text{s}}$ and the configurational entropy $S_{\text{c}}$ in these active liquids.  Our current work provides a thermodynamic way to understand the glassy dynamics of active systems. Our work can be used to study the physical systems such as Janus colloidal particles and the study of tracer particles in the bacterial suspension. 
\label{conc-section}

\begin{acknowledgments}
  We thank Ethayaraja Mani for discussions and comments on the manuscript. All simulations were done on the HPC-Physics cluster of our group and the AQUA super cluster of IIT Madras. Support from the core research grant SP20210716PHSERB008690 from SERB, Government of India, is gratefully acknowledged.

\end{acknowledgments}

\bibliography{manuscript}
\end{document}